\documentclass[fleqn,10pt]{SelfArx} 
\usepackage[english]{babel} 
\usepackage{lipsum} 
\usepackage{url} 
\usepackage{balance}

\usepackage{graphicx}   
\usepackage{subcaption} 
\usepackage{caption}    
\usepackage{acronym}    
\usepackage{float}      
\usepackage{hyperref}
\usepackage{mathptmx}
\usepackage{graphicx}
\usepackage{times}
\usepackage{textcomp}
\usepackage{gensymb} 
\usepackage{amsmath}
\usepackage{comment}
\usepackage{multirow}
\usepackage[framemethod=tikz]{mdframed}
\usepackage{graphicx}
\usepackage[labelfont=bf]{subcaption}
\usepackage{stfloats}

\usepackage[export]{adjustbox}
\usepackage{multicol}
\usepackage[utf8]{inputenc}
\usepackage[T1]{fontenc}
\usepackage{graphicx}
\usepackage{url}
\usepackage{float}
\usepackage{placeins}
\usepackage{framed}

\usepackage{cite}
\usepackage{amsmath,amssymb,amsfonts}
\usepackage{algorithmic}
\usepackage{textcomp}
\usepackage{xcolor}
\usepackage{acronym}
\usepackage[utf8]{inputenc}
\usepackage{graphicx,url}
\usepackage{multirow}
\usepackage{geometry}
\usepackage{stfloats}
\usepackage{afterpage}
\usepackage{subcaption} 
\usepackage{tcolorbox}
\definecolor{mycolor}{rgb}{0,0,0}
\makeatletter
\newcommand{\mybox}[1]{%
	\setbox0=\hbox{#1}%
	\setlength{\@tempdima}{\dimexpr\wd0+13pt}%
	\begin{tcolorbox}[colframe=mycolor,boxrule=0.5pt,arc=4pt,
		left=6pt,right=6pt,top=6pt,bottom=6pt,boxsep=0pt,width=\@tempdima]
		#1
	\end{tcolorbox}
}
\makeatother
\usepackage[resetlabels,labeled]{multibib}
\newcites{A}{Appendix}
\usepackage[num]{abntex2cite}
\citebrackets[]
\definecolor{color1}{RGB}{0,0,90} 
\definecolor{color2}{RGB}{0,20,20} 
\usepackage{hyperref} 
\hypersetup{hidelinks,colorlinks,breaklinks=true,urlcolor=color2,citecolor=color1,linkcolor=color1,bookmarksopen=false,pdftitle={Title},pdfauthor={Author}}
\JournalInfo{\color{gray}\normalsize\sffamily\bfseries Revista de Informática Teórica e Aplicada - RITA - ISSN 2175-2745\\ Vol.~XX, Num.~XX~(2018)~11-XX} 
\Archive{\mybox{RESEARCH ARTICLE}} 

\PaperTitle{Data Augmentation and Convolutional Network Architecture Influence on Distributed Learning} 

\ShortTitle{Short Title English} 

\Titulo{Influência do Aumento de Dados e da Arquitetura de Redes Convolucionais no Aprendizado Distribuído} 

\PaperVol{XX} 
\PaperNum{X} 
\PaperAno{YYYY} 

\Authors{Victor Forattini Jansen\textsuperscript{1}, Emanuel Teixeira Martins\textsuperscript{1},  Yasmin Souza Lima\textsuperscript{1}, Flávio {de Oliveira Silva}\textsuperscript{2}\textsuperscript{3}, Rodrigo Moreira\textsuperscript{1}, Larissa Ferreira {Rodrigues Moreira}\textsuperscript{1}*} 
\affiliation{\textsuperscript{1}\textit{Institute of Exacts and Technological Sciences, Federal University of Viçosa (UFV), Brazil}} 
\affiliation{\textsuperscript{2}\textit{Faculty of Computing (FACOM), Federal University of Uberlândia (UFU), Brazil}} 
\affiliation{\textsuperscript{3}\textit{Department of Informatics, University of Minho (UMINHO), Portugal}}
\affiliation{*\textbf{Corresponding authors}: \{victor.jansen, emanuel.martins, yasmin.lima, rodrigo, larissa.f.rodrigues\}@ufv.br,
flavio@di.uminho.pt}

\Keywords{Distributed Learning --- Rice Classification --- Data Augmentation --- CNN --- Factorial Design.}

\newcommand{\keywordname}{Keywords} 

\PalavrasChave{Aprendizado Distribuído — Classificação de Arroz — Aumento de Dados — CNN — Análise Fatorial.}

\Doi{http://dx.doi.org/10.22456/2175-2745.XXXX} 
\DateR{dd/mm/yyyy} 
\DateA{dd/mm/yyyy} 

\Abstract{Convolutional Neural Networks (CNNs) have proven to be highly effective in solving a broad spectrum of computer vision tasks, such as classification, identification, and segmentation. These methods can be deployed in both centralized and distributed environments, depending on the computational demands of the task. While much of the literature has focused on the explainability of CNNs, which is essential for building trust and confidence in their predictions, there remains a gap in understanding their impact on computational resources, particularly in distributed training contexts. In this study, we analyze how CNN architectures primarily influence model accuracy and investigate additional factors that affect computational efficiency in distributed systems. Our findings contribute valuable insights for optimizing the deployment of CNNs in resource-intensive scenarios, paving the way for further exploration of variables critical to distributed learning.}

\Resumo{As Redes Neurais Convolucionais (CNNs) têm se mostrado altamente eficazes na solução de uma ampla gama de tarefas de visão computacional, como classificação, identificação e segmentação. Esses métodos podem ser implementados tanto em ambientes centralizados quanto distribuídos, dependendo das exigências computacionais da tarefa. Embora grande parte da literatura tenha se concentrado na explicabilidade das CNNs, o que é essencial para gerar confiança em suas previsões, ainda há uma lacuna no entendimento de seu impacto nos recursos computacionais, especialmente em contextos de treinamento distribuído. Neste estudo, analisamos como as arquiteturas de CNN influenciam principalmente a precisão do modelo e investigamos fatores adicionais que afetam a eficiência computacional em sistemas distribuídos. Nossos resultados oferecem insights valiosos para otimizar a implementação de CNNs em cenários com alta demanda de recursos, abrindo caminho para a exploração de variáveis críticas para o aprendizado distribuído.}

\begin{document}
\acrodef{AI}{Artificial Intelligence}
\acrodef{AIaaS}{Artificial Intelligence as a Service}
\acrodef{Ablation-CAM}{Ablation-based Class Activation Mapping}
\acrodef{API}{Application Programming Interface}
\acrodef{ANOVA}{Analysis of Variance}
\acrodef{CNN}{Convolutional Neural Network}
\acrodef{CNNs}{Convolutional Neural Networks}
\acrodef{CPU}{Central Processing Unit}
\acrodef{DQN}{Deep Q-Network}
\acrodef{DA}{Data Augumentation}
\acrodef{DT}{Decision Tree}
\acrodef{FTP}{File Transfer Protocol}
\acrodef{FPGA}{Field Programmable Gate Array}
\acrodef{FAO}{Food and Agriculture Organization}
\acrodef{GPU}{Graphics Processing Unit}
\acrodef{GPUs}{Graphics Processing Units}
\acrodef{GB}{Gradient Boost}
\acrodef{HPS}{High Performance Server}
\acrodef{HPC}{High Performance Computing}
\acrodef{KNN}{K-Nearest Neighbors}
\acrodef{LAN}{Local Area Network}
\acrodef{MLaaS}{Machine Learning as a Service}
\acrodef{MLOps}{Machine Learning Operations}
\acrodef{MLP}{Multilayer Perceptron}
\acrodef{MSE}{Mean Squared Error}
\acrodef{ML}{Machine Learning}
\acrodef{NLP}{Natural Language Processing}
\acrodef{NPDT}{Professional Dental Treatment}
\acrodef{P2P}{Peer-to-Peer}
\acrodef{PAIaaS}{Pervasive Artificial Intelligence as a Service}
\acrodef{ROI}{Region of Interest}
\acrodef{RF}{Random Forest}
\acrodef{RL}{Reinforcement Learning}
\acrodef{PoC}{Proof of Concept}
\acrodef{SSD}{Single-Shot Multibox Detector}
\acrodef{SGD}{Stochastic Gradient Descent}
\acrodef{SLA}{Service Level Agreement}
\acrodef{SVM}{Support Vector Machine}
\acrodef{TOR}{The Onion Router}
\acrodef{VoIP}{Voice over IP}
\acrodef{VPN}{Virtual Private Network}
\acrodef{YOLOv3}{You Only Look Once version 3}

\setcounter{page}{11}

\flushbottom 
\maketitle 
\thispagestyle{empty} 


\section{Introduction} 

\addcontentsline{toc}{section}{Introduction} 

The increasing complexity of deep learning models, particularly Convolutional Neural Networks (CNNs), has heightened the demand for optimizing distributed training methods. Distributed training is essential for handling large datasets and deep architectures efficiently, enabling better utilization of computational resources across multiple machines. This study focuses on evaluating how key factors — such as network depth and data augmentation (DA) — affect both performance and resource utilization in distributed environments~\cite{Song2024}.

Managing computational resources efficiently is critical during the training phase, as both time and resources are valuable, especially with complex architectures. The selection of a CNN architecture has a direct influence on hardware usage metrics~\cite{RodriguesMoreira2023}. For example, the number of layers and the complexity of numerical computations in the network significantly affect GPU, CPU usage, memory consumption, and network traffic~\cite{wiedemann2020transactions}. However, the specific impact of these factors in a distributed setting has not been fully explored. Despite there are efforts towards employed \ac{CNN} to enrich disease classification on images, there is a lack of studies which dive into \ac{CNN} hardware demands footprint.

This paper examines the impact of factors such as \ac{CNN} architecture and the introduction of \ac{DA} on various response variables, including \ac{GPU}, \ac{CPU} usage, memory consumption, network packets, and accuracy. To further explore these relationships, we conducted experiments based on a factorial design within a distributed training environment, offering novel insights into how \ac{CNN} architectures and \ac{DA} influence the aforementioned metrics -- an aspect not yet extensively explored in the state of the art. While our experiments use a rice disease classification dataset as a case study -- because of the quality of the images and the diseases classification, the primary objective is to understanding the broader impact of \ac{CNN} architecture and \ac{DA} in distributed training scenarios~\cite{Petchiammal2023}. 

The remainder of this paper is organized as follows. Section~\ref{sec:related_work} surveys the related work. Section~\ref{sec:method} presents our proposed approach to explore data augmentation and deep learning architecture influence on distributed learning. Section~\ref{sec:case_study} describes our results and discusses the main findings. Finally, Section~\ref{sec:conclusion} presents the conclusion and opportunities for future work.


\section{Related Work}\label{sec:related_work}

In recent years, significant advances have been made worldwide in applying deep learning and computer vision techniques to security\cite{Costa2021} and health~\cite{daRocha2020, Bouzon2023} agricultural crops~\cite{moreira2022agrolens}, particularly for classifying diseases in plant leaves. This section reviews recent advancements in computer vision techniques to address this challenge by focusing on the impacting resources.  In the state-of-the-art, there are limited approaches that measure the impact that \ac{CNNs} and \ac{DA} techniques impose on the underlying hardware, particularly from the point of view of network consumption~\cite{Park2020, Mohaidat2024}.

Petchiammal et al. \cite{Petchiammal2023} developed the Paddy Doctor dataset, which contains 16.225 classified images across 12 disease labels and one healthy label. The photos were collected from real paddy fields, annotated under professional agricultural supervision, and used as a benchmark in different CNN models, such as VGG16, MobileNet, Xception, and Resnet34, with data augmentation (rotation, shear intensity, zoom, width, and height shift, and horizontal flip). In our study, we use this pre-classified dataset to evaluate the impacts of distributed learning on the classification of rice leaf disease.

Aggarwal et al. \cite{Aggarwal2023Federated} proposed a federated transfer learning (F-TL) framework to address the challenges of rice-leaf disease classification, particularly in scenarios where data privacy and distributed data sources are concerned. The study presents an unbalanced dataset with four diseases that were separated on both IID (independent and identically distributed) and non-IID (non-identically distributed) and trained in models such as EfficientNEtB3 and MobileNetV2, demonstrating strong performance across IID and non-IID datasets.

Ni et. al \cite{Ni2023} introduced an improved model based on the RepVGG architecture, integrated with the Efficient Channel Attention (ECA) mechanism. The proposed model, RepVGG-ECA, is designed to improve classification through the attention mechanism. The study used the Paddy Doctor dataset under data augmentation such as inversion, saturation modification, contrast, and adding blur to focus on the diseases. Comparing the accuracy, macro-f1, macro-precision and macro-recall, they found out that their introduced model achieved 97.06\% accuracy, outperforming other models such as ResNet34 and ShuffleNetV2.

Yang et al. \cite{Yang2023Lightweight} proposed the DGLNet, a lightweight network specifically designed for identifying rice diseases. Their network integrates a Global Attention Module (GAM) - which is engineered to capture critical information in complex and/or noisy environments - and a Dynamic Representation Module (DRM), that is designed to improve feature representation through a self-developed for-dimensional flexible convolution (4D-FConv). The study used two datasets, including the Paddy Doctor, achieving 99.71\% of accuracy.

Senthy et al. \cite{senthy2020svm} compared the performance of 11 deep CNN and 2 shallow CNN models with different architectures for disease detection. In study, deep features extracted from these models were combined with transfer learning and an SVM for classification tasks. The ResNet50 model, when paired with SVM, achieved the highest accuracy and F1 score of 98.38\%. Similarly, the SVM classifier, when using features obtained from the AlexNet CNN, reached an accuracy of 96.8\%.
When analyzing shallow CNN architecture as MobileNetV2, the results was comparable to ResNet50, achieving 97.96\% of accuracy. 

While previous works have primarily focused on evaluating models to enhance classification through computer vision, many have not sufficiently explored how these models impact hardware usage. To fill this gap, our paper studies the interplay between network depth and distributed training, analyzing hardware and training metrics to determine how the CNN architecture and data augmentation impact the process.


\section{Influence Assessment Method}\label{sec:method}
In this study, we employed a \(2^2\) factorial design, a specific type of \ac{ANOVA}, to systematically investigate the effects of two independent factors on the performance of \ac{CNN} in a classification task. According to Table~\ref{tab:experiment_rationale}, the factors considered were \ac{DA} and \ac{CNN} Architectures, each evaluated at two levels: with-\ac{DA} ($1$) and without-\ac{DA} ($-1$) for the Data Augmentation factor, and shallow-\ac{CNN} ($1$) and deep-\ac{CNN} ($-1$) for the \ac{CNN} Architectures factor. The factorial design allows for the examination of the main effects of each factor and the interaction effect between the factors. 

\begin{table}[!ht]
\caption{Factorial Design.}
\renewcommand{\arraystretch}{1.3}
\label{tab:experiment_rationale}
\resizebox{\columnwidth}{!}{%
\begin{tabular}{ll|ccccc|}
\cline{3-7}
                                                                                                        &                                      & \multicolumn{5}{c|}{\textbf{Response Variable}}                                                                                                                                                                                                                                                             \\ \hline
\multicolumn{1}{|c|}{\textbf{Factors}}                                                                  & \multicolumn{1}{c|}{\textbf{Levels}} & \multicolumn{1}{c|}{\textbf{GPU}}           & \multicolumn{1}{c|}{\textbf{\begin{tabular}[c]{@{}c@{}}Network \\ Packets\end{tabular}}} & \multicolumn{1}{c|}{\textbf{CPU}}           & \multicolumn{1}{c|}{\textbf{\begin{tabular}[c]{@{}c@{}}Memory \\ Consumption\end{tabular}}} & \textbf{Accuracy}      \\ \hline
\multicolumn{1}{|l|}{\multirow{2}{*}{\begin{tabular}[c]{@{}l@{}}\textbf{A:} Data \\ Augumentation\end{tabular}}} & with-DA (1)                          & \multicolumn{1}{c|}{\multirow{4}{*}{$y_1$}} & \multicolumn{1}{c|}{\multirow{4}{*}{$y_2$}}                                              & \multicolumn{1}{c|}{\multirow{4}{*}{$y_3$}} & \multicolumn{1}{c|}{\multirow{4}{*}{$y_4$}}                                                 & \multirow{4}{*}{$y_5$} \\ \cline{2-2}
\multicolumn{1}{|l|}{}                                                                                  & without-DA (-1)                      & \multicolumn{1}{c|}{}                       & \multicolumn{1}{c|}{}                                                                    & \multicolumn{1}{c|}{}                       & \multicolumn{1}{c|}{}                                                                       &                        \\ \cline{1-2}
\multicolumn{1}{|l|}{\multirow{2}{*}{\begin{tabular}[c]{@{}l@{}}\textbf{B:} CNN \\ Architecture\end{tabular}}}   & shallow-CNN (1)                      & \multicolumn{1}{c|}{}                       & \multicolumn{1}{c|}{}                                                                    & \multicolumn{1}{c|}{}                       & \multicolumn{1}{c|}{}                                                                       &                        \\ \cline{2-2}
\multicolumn{1}{|l|}{}                                                                                  & deep-CNN (-1)                        & \multicolumn{1}{c|}{}                       & \multicolumn{1}{c|}{}                                                                    & \multicolumn{1}{c|}{}                       & \multicolumn{1}{c|}{}                                                                       &                        \\ \hline
\end{tabular}%
}
\end{table}

This study aims to comprehend the scientific basis for the impact of two key elements, \ac{DA} and \ac{CNN} architecture, on the hardware used in distributed learning, particularly in quantitative analysis and the interplay between these two components. Given the above problem statement, we formulated the factorial experiment combinations according to Table~\ref{tab:experiment_rationale}. In our experimental design, we identified six (6) response variables that enable us to assess the impact of hardware on distributed learning environments as in Table~\ref{tab:experiment_rationale}. For shallow-\ac{CNN}, we refer to those with few convolutional layers, whereas deep-\ac{CNN} refers to a larger number of layers.

Mathematically, the experiment can be modeled using the following linear model:
\[
Y = \mu + \tau_A + \tau_B + \tau_{AB} + \epsilon
\]
where \(Y\) represents the observed outcome (e.g., accuracy or F1-Score), \(\mu\) is the overall mean, \(\tau_A\) and \(\tau_B\) are the main effects of \ac{DA} and \ac{CNN} Architectures, respectively, \(\tau_{AB}\) is the interaction effect between the two factors, and \(\epsilon\) represents the random error term. The main effects and interaction influence are estimated using the following contrasts:
\[
\tau_A = \frac{1}{2} \left[ (Y_{A_1B_1} + Y_{A_1B_0}) - (Y_{A_0B_1} + Y_{A_0B_0}) \right]
\]
\[
\tau_B = \frac{1}{2} \left[ (Y_{A_1B_1} + Y_{A_0B_1}) - (Y_{A_1B_0} + Y_{A_0B_0}) \right]
\]
\[
\tau_{AB} = \frac{1}{2} \left[ (Y_{A_1B_1} - Y_{A_1B_0}) - (Y_{A_0B_1} - Y_{A_0B_0}) \right]
\]

\begin{figure}[!ht]
  \includegraphics[width=1\columnwidth]{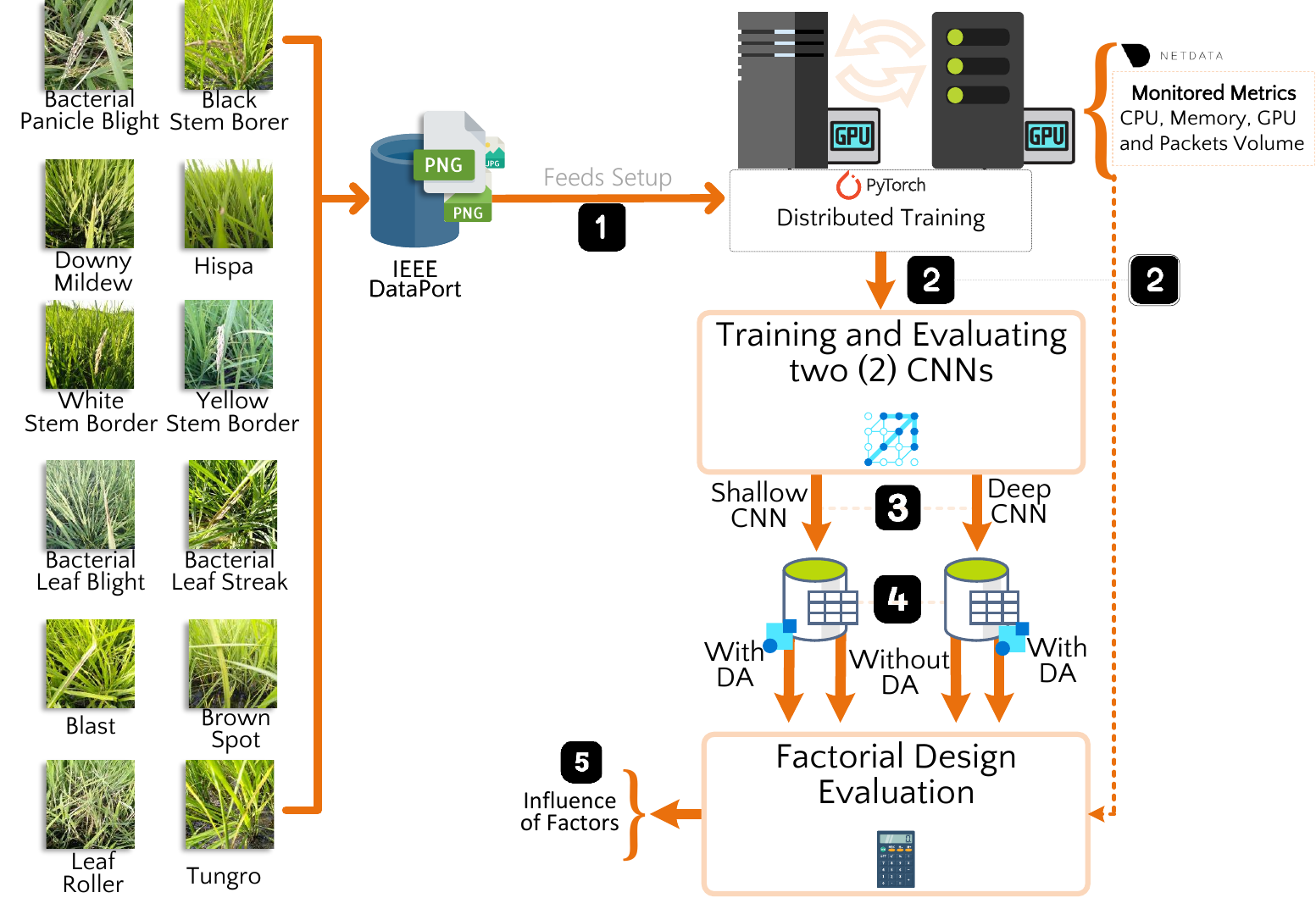}
  \caption{Proposed method.}
  \label{fig:proposed_method}
\end{figure}

Our design provides a comprehensive understanding of how the different levels of \ac{DA} and \ac{CNN} Architectures influence the distributed learning hardware, individually and in combination. The interaction term, in particular, reveals whether the effect of one factor depends on the level of the other factor. In Fig.~\ref{fig:proposed_method}, we present our method to measure our factor influence on hardware in a distributed training setup.

The Paddy Doctor dataset, depicted in left side of Fig.~\ref{fig:proposed_method}, is a visual image dataset designed for automated paddy disease classification. It comprises 16,225 annotated images of paddy leaves, categorized into 13 classes: 12 diseases (e.g., Bacterial Leaf Blight, Brown Spot, Tungro) and normal leaves. Collected from real paddy fields and annotated by an agricultural officer, this dataset is valuable for developing and benchmarking deep learning models for paddy disease detection~\cite{Petchiammal2023}.

According to Fig.~\ref{fig:proposed_method}, step one (1) of our method is to feed our distributed training environment with dataset images. In the step two (2), we have parallel procedures flow, while our distributed testbed carried out traning over data, we measured and collect resouces metrics to measure the the impact. We employed the monitoring tool NetData to record the hardware consumption of the \ac{CPU}, memory, network packets, and GPU. Step three (3) refers to the experiment factors combinations with shallow-\ac{CNN} (MobileNetV2-100 with bath normalization) and deep-\ac{CNN} (MobileOne-S1). Step four (4) refers to the level combinations regarding the use of \ac{DA} or not. 

Our employed data augmentation pipeline (step 4 of Fig.~\ref{fig:proposed_method}) includes random rotations ($\pm5^\circ$), affine transformations with shear (0.2) and slight translations ($\pm5\%$), random resized cropping (80-100\%) of the original image size), horizontal flipping, and color jittering. These enhancements may improve model generalization by introducing variability in spatial orientation, scale, color, and positioning while preserving the essential features of the images. In step five (5) we carried out \ac{ANOVA} to asses the factors and levels on different variables of response.

\section{Evaluation and Discussion}\label{sec:case_study}

We conducted experiments using the hardware specifications listed in Table~\ref{tab:experimental_testbed} to measure the factors influencing hardware resources in a distributed training scenario on top of Ubuntu 20.04 Long Term Evolution (LTE). Specifically, our distributed learning environment consists of two servers exchanging training gradients and \ac{CNN} weights over a \ac{LAN} with a 1 Gbps interface, which was negotiated to operate at 100 Mbps. We employed the Torch Distributed Data Parallel as the backend of distributed training. 

\begin{table}[!ht]
\caption{Experimental Testbed.}
\label{tab:experimental_testbed}
\renewcommand{\arraystretch}{1.2}
\resizebox{\columnwidth}{!}{%
\begin{tabular}{|c|c|c|c|c|}
\hline
\textbf{Server} & \textbf{CPU}                            & \textbf{RAM} & \textbf{GPU}             & \textbf{Ethernet} \\ \hline
\textbf{\#1}    & \begin{tabular}[c]{@{}c@{}}Intel(R) Core(TM) \\ i5-4430 3.00GHz\end{tabular} & 32 GB        & \begin{tabular}[c]{@{}c@{}}GeForce  \\ RTX 4060 Ti 8 GB\end{tabular}  & 1Gbps             \\ \hline
\textbf{\#2}    & \begin{tabular}[c]{@{}c@{}}Intel(R) Core(TM) \\ i5-4430 CPU 3.00GHz\end{tabular} & 16 GB        & \begin{tabular}[c]{@{}c@{}}GeForce \\ GTX 1050 Ti 4 GB\end{tabular} & 1Gbps             \\ \hline
\end{tabular}%
}
\end{table}

\begin{figure*}[!ht]
    \centering
    \begin{tabular}{ccc}
			\includegraphics[width=0.28\textwidth]{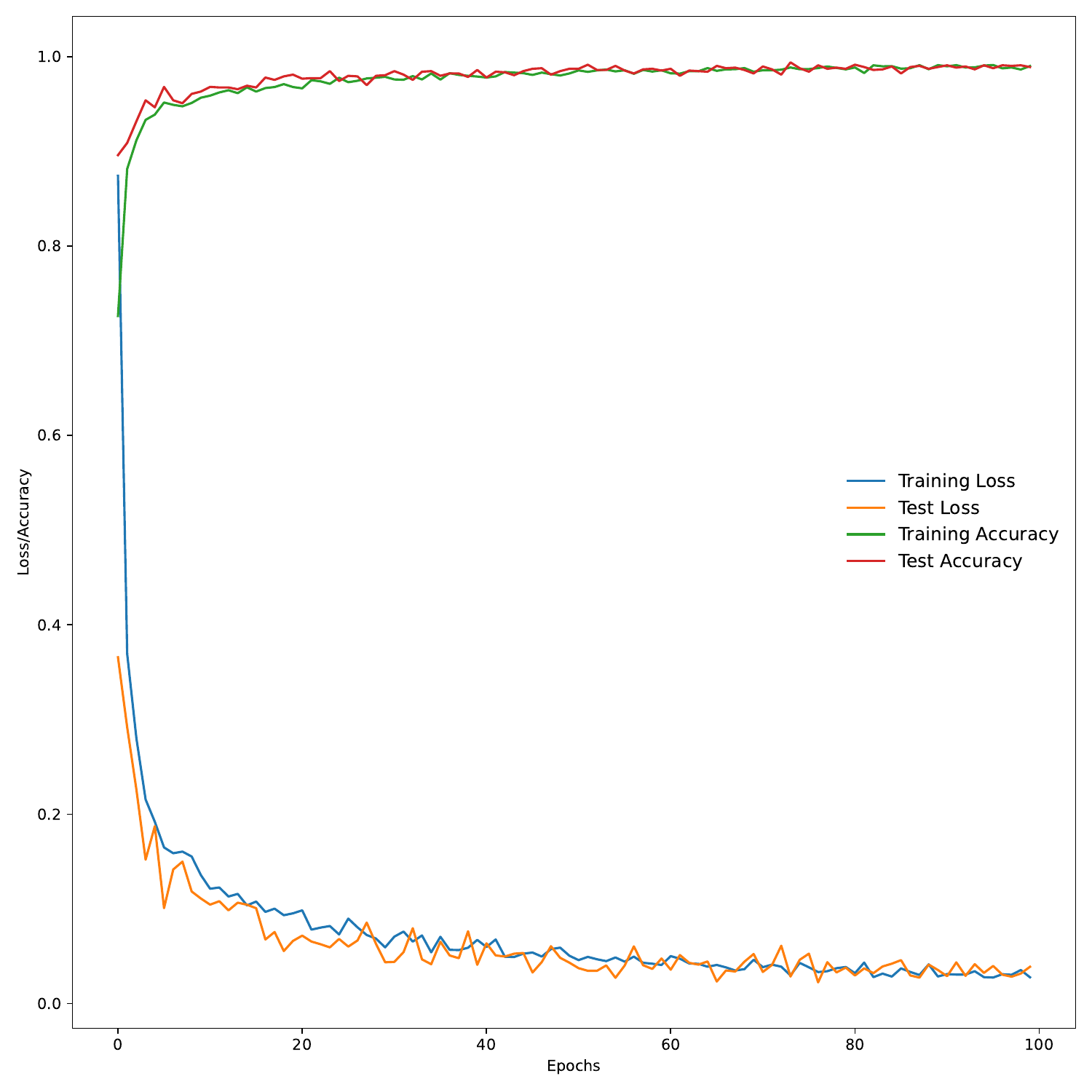} &
			\includegraphics[width=0.28\textwidth]{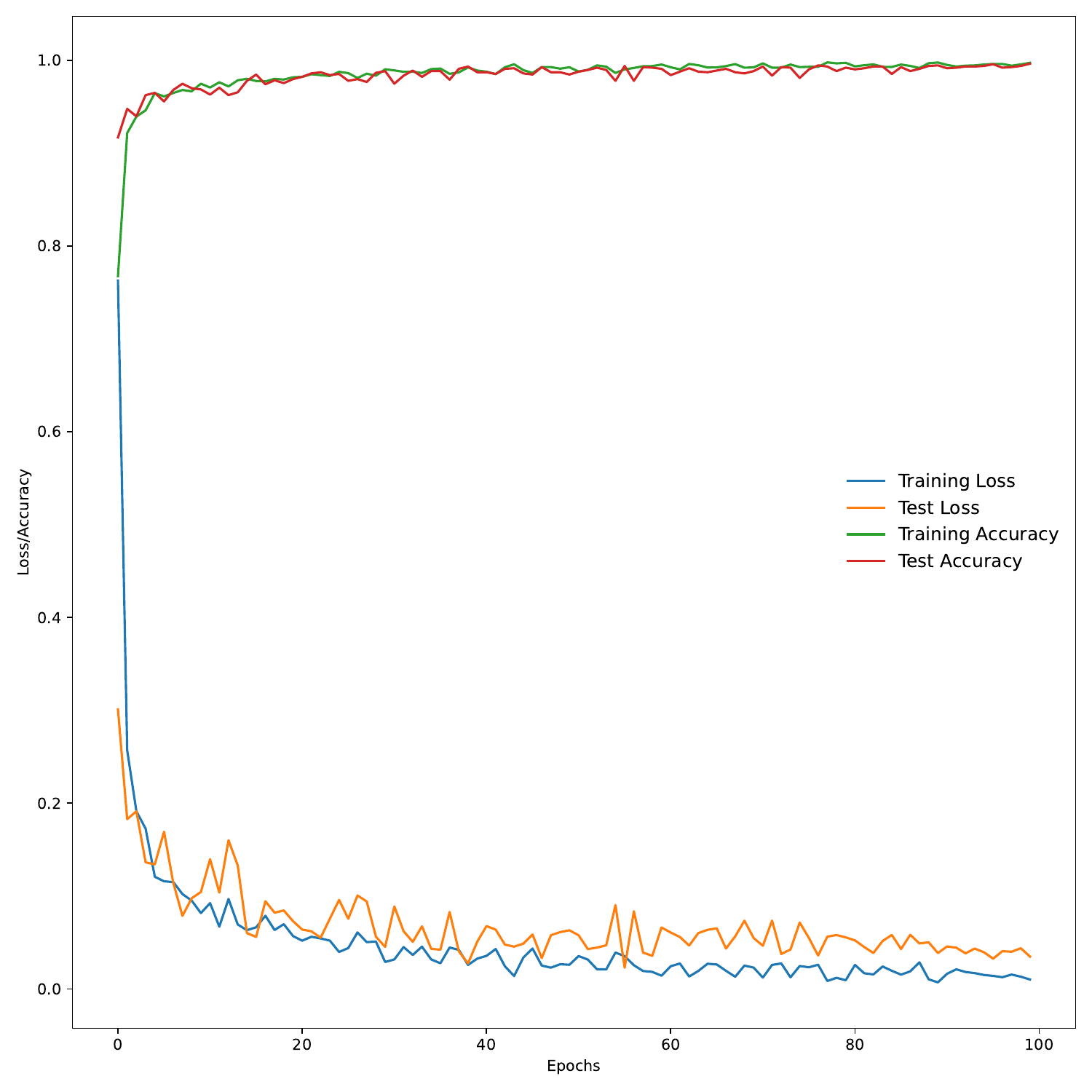} &
   			\includegraphics[width=0.28\textwidth]{37_shallow_non-da_training_testing_metrics.pdf} \\
                (a) with-\ac{DA} and shallow-\ac{CNN}. & (b) without-\ac{DA} and shallow-\ac{CNN}.& (c) with-\ac{DA} and deep-\ac{CNN}.\\		      
                \includegraphics[width=0.28\textwidth]{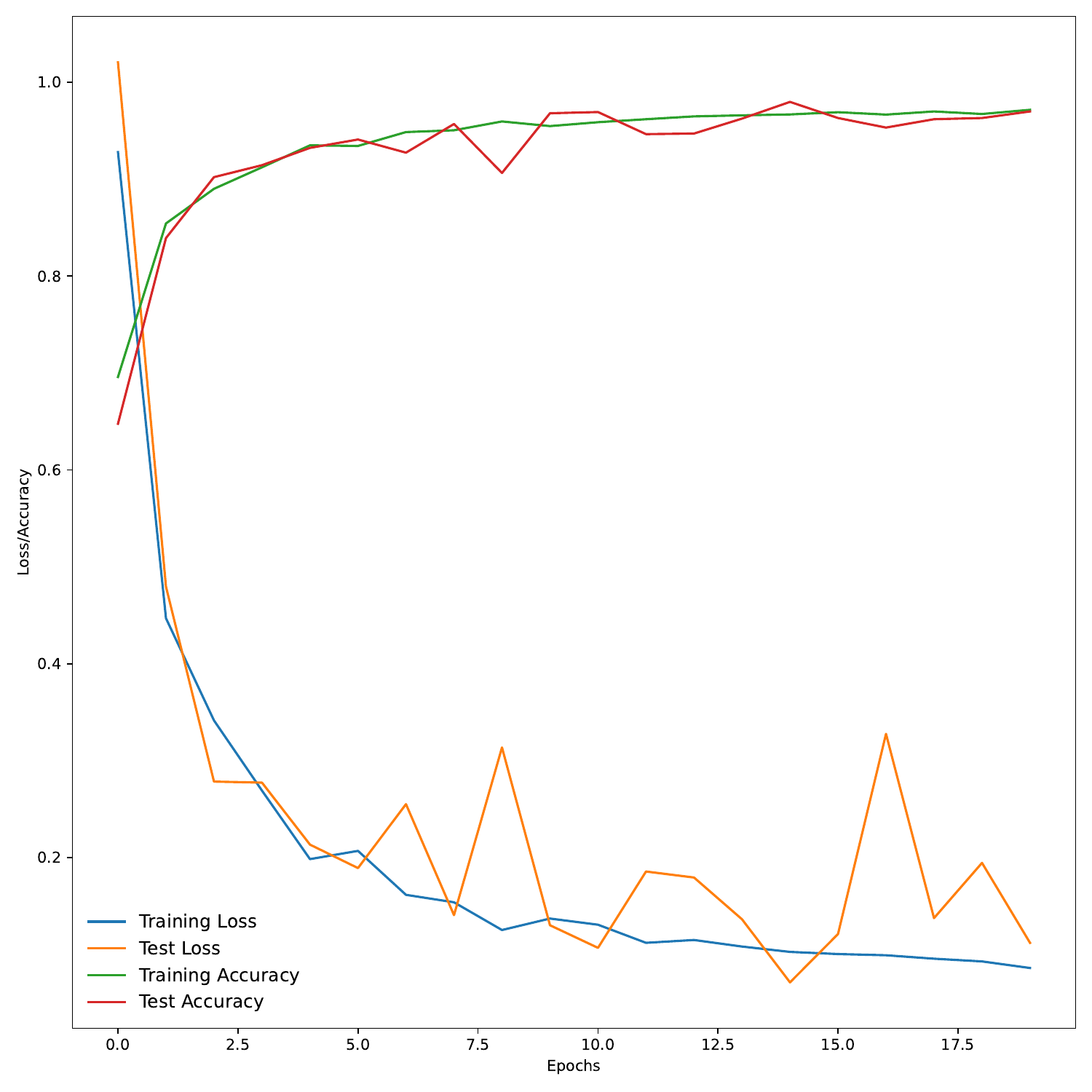} & 
			\includegraphics[width=0.28\textwidth]{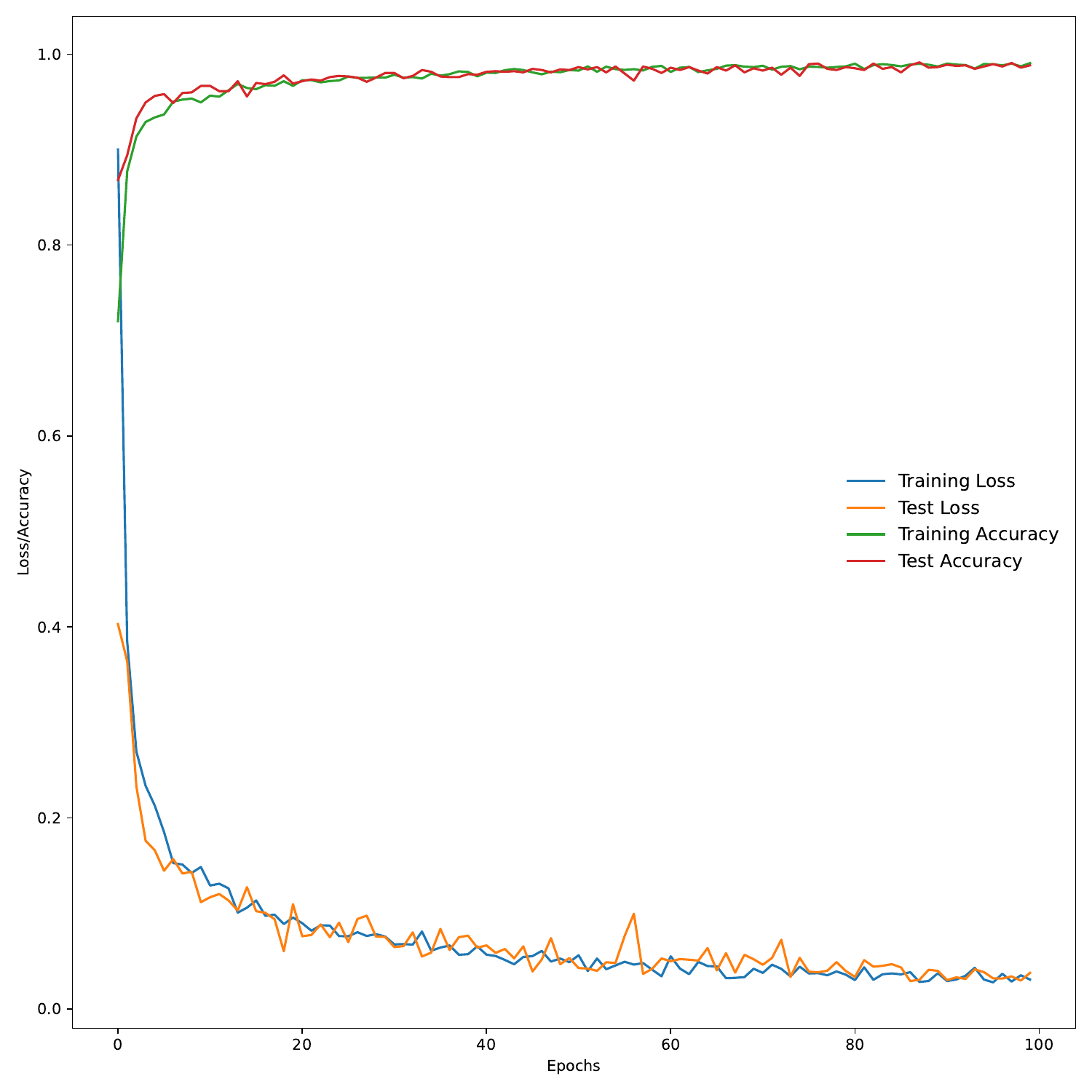} &
   			\includegraphics[width=0.28\textwidth]{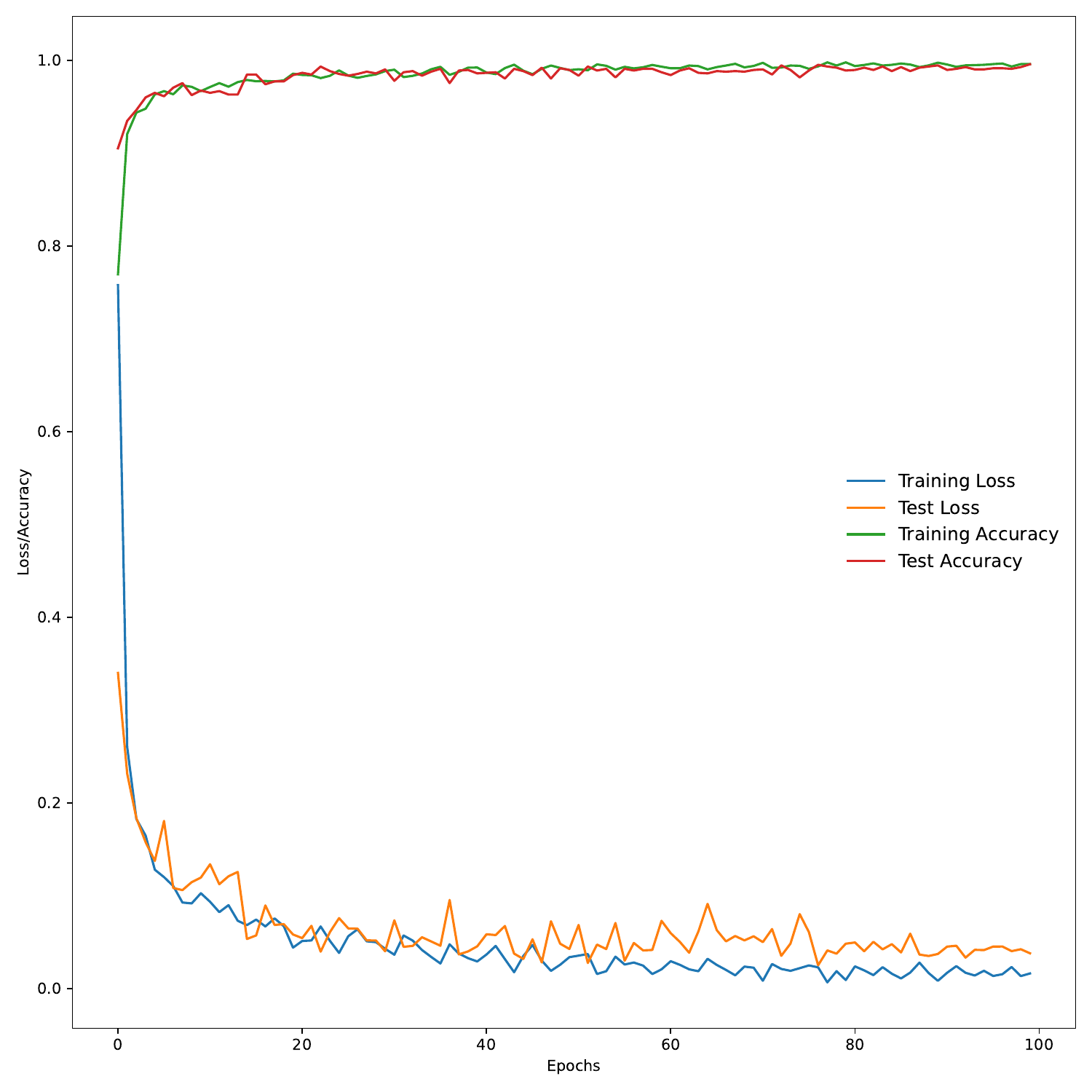} \\                
			(d) without-\ac{DA} and deep-\ac{CNN}. & (e) with-\ac{DA} and shallow-\ac{CNN}. & (f) without-\ac{DA} and shallow-\ac{CNN}.\\	  
    \end{tabular}

    \begin{tabular}{cc}

             \includegraphics[width=0.28\textwidth]{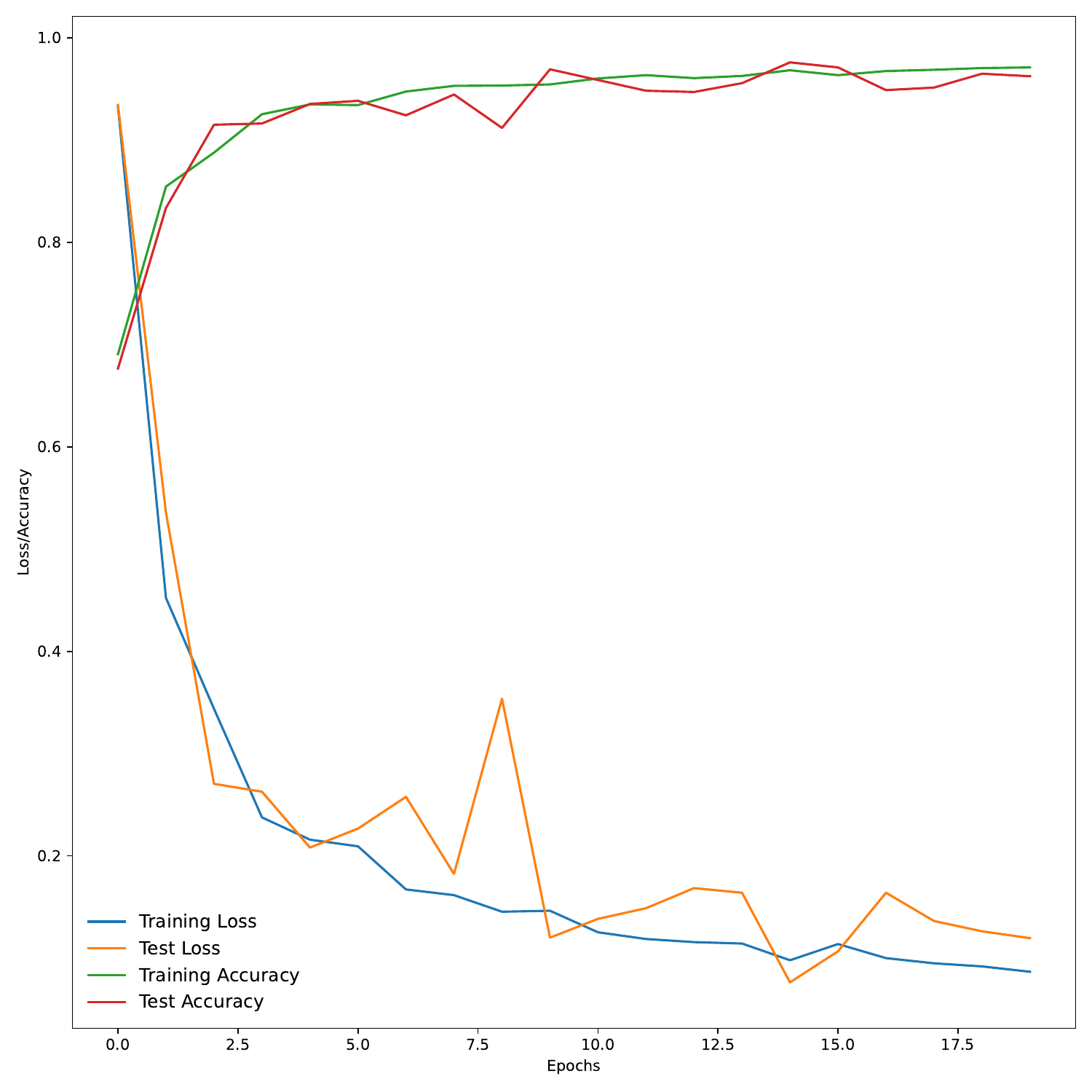} &
		   \includegraphics[width=0.28\textwidth]{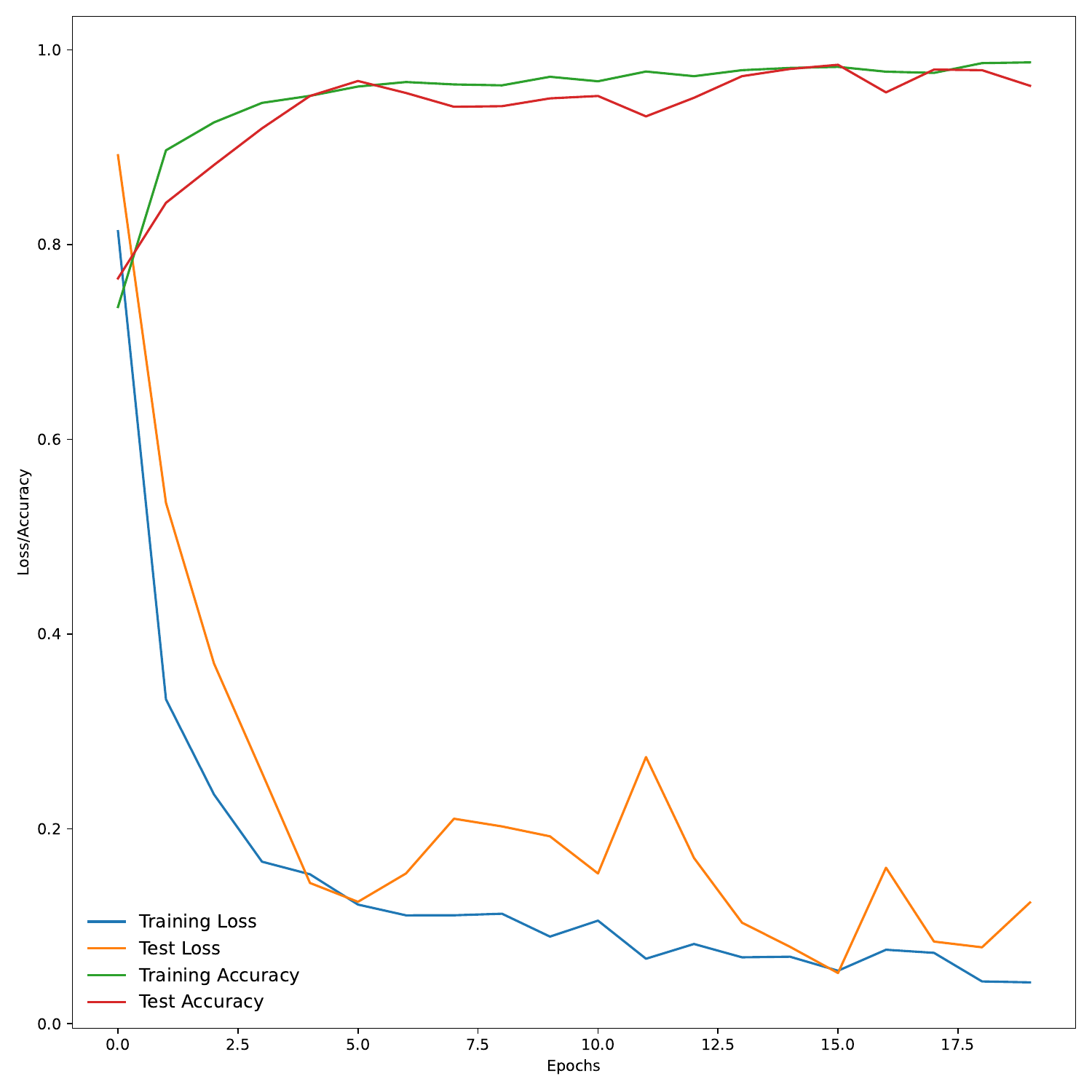} \\   
            (g) with-\ac{DA} and deep-\ac{CNN}. & (h) without-\ac{DA} and deep-\ac{CNN}.\\		    
    \end{tabular}
    \caption{Training and Testing Metrics for Server \#1 (a), (b), (c), (d) and Server \#2 (e), (f), (g), (h).}
   \label{fig:metrics_comparison_server}
\end{figure*}

Considering the training and validation performance, we computed the accuracy and loss graphs for both the training and testing phases in our distributed setup. As shown in Fig.\ref{fig:metrics_comparison_server} (a), (b), (c), and (d), empirical evidence suggests that for all experiment combinations, there was learning progress over the epochs in our proposed evaluation of the impact of factors such as \ac{DA} and \ac{CNN} architecture, specifically on Server \#1. Although we standardized the number of epochs to 100 in our distributed scenario, for graphical presentation, we applied early stopping in the training process, as shown in Fig.\ref{fig:metrics_comparison_server}(c), and \ref{fig:metrics_comparison_server}(d), concluding at epoch 20 due to the absence of further improvements.

Similarly, we assessed the behavioral performance of the model trained on Server \#2. Empirical evidence also showed that the model used in our performance evaluation was able to learn correctly over epochs. This is evident in Fig.\ref{fig:metrics_comparison_server} (e), (f), (g), and (h), where all four experimental combinations showed a decrease in loss with exponential decay and a sigmoid increase in accuracy.

Regarding the central objective of our evaluation, we conducted a factorial analysis based on the \ac{ANOVA} method considering the experimental variations in Table~\ref{tab:experiment_rationale}. The results of the experiments are reported in Table~\ref{tab:factorial_average_measurements}, showing the average measurements for each response variable. We assessed the average (\%) consumption of $Y_{GPU}$, $Y_{NetworkPackets}$, $Y_{CPU}$, $Y_{Memory}$, and $Y_{Accuracy}$ for the four experimental combinations. Factors such as $Y_{Memory}$, although there were experimental variations, maintained its average consumption.

\begin{table}[!ht]
\centering
\renewcommand{\arraystretch}{1.1}
\caption{Average of Response Variable in Factorial Experiments.}
\label{tab:factorial_average_measurements}
\resizebox{\columnwidth}{!}{%
\begin{tabular}{l|ccccc|}
\cline{2-6}
                                                                                                      & \multicolumn{5}{c|}{\textbf{\begin{tabular}[c]{@{}c@{}}Response \\ Variable\end{tabular}}}                                                                                                                                                                                    \\ \hline
\multicolumn{1}{|c|}{\textbf{Experiment}}                                                             & \multicolumn{1}{c|}{\textbf{$Y_{GPU}$ (\%)}} & \multicolumn{1}{c|}{\textbf{\begin{tabular}[c]{@{}c@{}}$Y_{Network Packets}$ \\ (Pkts/s)\end{tabular}}} & \multicolumn{1}{c|}{\textbf{\begin{tabular}[c]{@{}c@{}}$Y_{CPU}$ \\ (\%)\end{tabular}}} & \multicolumn{1}{c|}{\textbf{\begin{tabular}[c]{@{}c@{}}$Y_{Memory}$ \\ (\%)\end{tabular}}} & \multicolumn{1}{c|}{\textbf{\begin{tabular}[c]{@{}c@{}}$Y_{Accuracy}$ \\ (\%)\end{tabular}}} \\ \hline
\multicolumn{1}{|l|}{\begin{tabular}[c]{@{}l@{}}with-DA (1)\\  and shallow-CNN (1)\end{tabular}}      & \multicolumn{1}{c|}{95,12}                & \multicolumn{1}{c|}{19994,50}                                                                              & \multicolumn{1}{c|}{51,15}                & \multicolumn{1}{c|}{81,70}                   & 98,71                     \\ \hline
\multicolumn{1}{|l|}{\begin{tabular}[c]{@{}l@{}}without-DA (-1) \\  and shallow-CNN (1)\end{tabular}} & \multicolumn{1}{c|}{97,18}                & \multicolumn{1}{c|}{15698,97}                                                                              & \multicolumn{1}{c|}{47,38}                & \multicolumn{1}{c|}{81,75}                   & 99,60                     \\ \hline
\multicolumn{1}{|l|}{\begin{tabular}[c]{@{}l@{}}with-DA (1)\\  and deep-CNN (-1)\end{tabular}}        & \multicolumn{1}{c|}{97,21}                & \multicolumn{1}{c|}{19973,00}                                                                              & \multicolumn{1}{c|}{47,03}                & \multicolumn{1}{c|}{80,45}                   & 94,09                     \\ \hline
\multicolumn{1}{|l|}{\begin{tabular}[c]{@{}l@{}}without-DA (-1)\\  and deep-CNN (-1)\end{tabular}}    & \multicolumn{1}{c|}{98,29}                & \multicolumn{1}{c|}{10526,36}                                                                              & \multicolumn{1}{c|}{43,85}                & \multicolumn{1}{c|}{81,45}                   & 96,58                     \\ \hline
\end{tabular}%
}
\end{table}

\begin{figure*}[!ht]
\centering
  \includegraphics[width=0.45\textwidth]{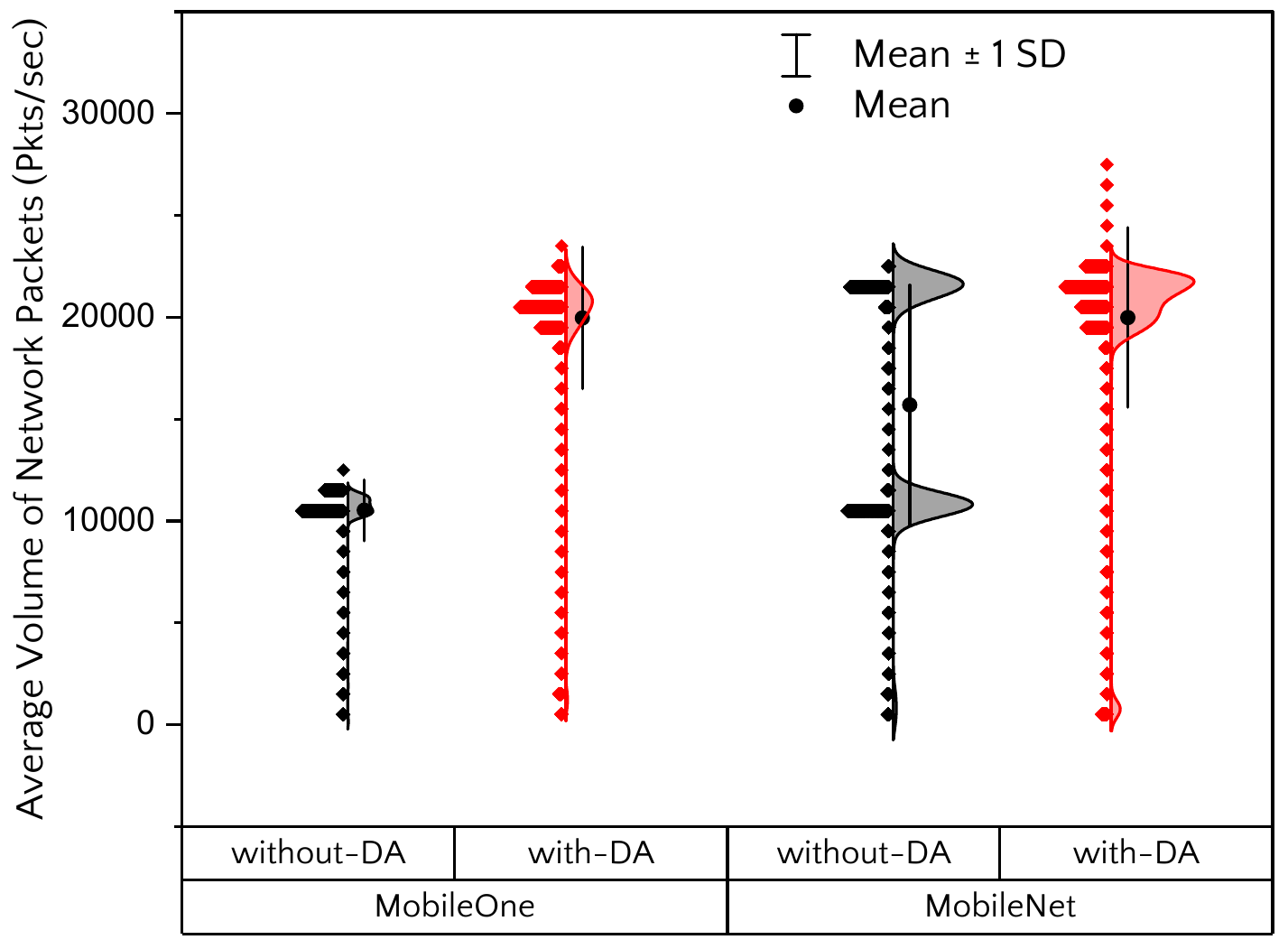}
  \caption{\ac{DA} is affecting the distribution of average packet volume.}
  \label{fig:violing_networkpacket_distribution}
\end{figure*}

On the other hand, the response variable $Y_{NetworkPackets}$ exhibited non-uniform behavior. We found that the experimental combination with-\ac{DA} imposes a higher volume of network packets in distributed training. Fig.~\ref{fig:violing_networkpacket_distribution} shows the grouped violin plot of two categorical variables: with-\ac{DA} and without-\ac{DA}. As observed, for both neural network architectures (deep and shallow), the introduction of \ac{DA} results in a higher volume of network packets, as indicated by the accumulation of instances in the plot.

As shown in Table~\ref{tab:label_2_here} we advanced in our analysis by evaluating the influence of the factors \ac{DA} ($\tau_A$) and \ac{CNN} architecture ($\tau_B$) both individually and in combination ($\tau_{AB}$) on the response variables. The factor $\tau_{B}$ (\ac{CNN} architecture) has an influence of $48.64\%$ on the response variable $Y_{GPU}$, followed by an influence of $46.83\%$ from the factor $\tau_{A}$. Regarding the response variable $Y_{NetworkPackets}$, we validated the initial hypothesis regarding the influence of introducing \ac{DA} in training. We found that introducing \ac{DA} in distributed learning results in a $77.92\%$ influence on the volume of packets transmitted during deployment.

\begin{table}[!ht]
\centering
\caption{Comparison of the influence of factors on different response variables.}
\renewcommand{\arraystretch}{1.3}
\label{tab:label_2_here}
\resizebox{\columnwidth}{!}{%
\begin{tabular}{c|ccccc|}
\cline{2-6}
\multicolumn{1}{l|}{}                                                                       & \multicolumn{5}{c|}{\textbf{Influence Measurement (\%)}}                                                                                                                                        \\ \hline
\multicolumn{1}{|c|}{\textbf{\begin{tabular}[c]{@{}c@{}}Factors \\ Iteration\end{tabular}}} & \multicolumn{1}{c|}{\textbf{$Y_{GPU}$}} & \multicolumn{1}{c|}{\textbf{$Y_{Network Packets}$}} & \multicolumn{1}{c|}{\textbf{$Y_{CPU}$}} & \multicolumn{1}{c|}{\textbf{$Y_{Memory}$}} & \textbf{$Y_{Accuracy}$} \\ \hline
\multicolumn{1}{|c|}{$\tau_A$}                                                                    & \multicolumn{1}{c|}{46,83}           & \multicolumn{1}{c|}{77,92}                       & \multicolumn{1}{c|}{45,07}           & \multicolumn{1}{c|}{24,53}              & 15,86                \\ \hline
\multicolumn{1}{|c|}{$\tau_B$}                                                                    & \multicolumn{1}{c|}{48,64}           & \multicolumn{1}{c|}{11,13}                        & \multicolumn{1}{c|}{54,61}           & \multicolumn{1}{c|}{54,75}              & 80,60                \\ \hline
\multicolumn{1}{|c|}{$\tau_{AB}$}                                                                   & \multicolumn{1}{c|}{4,53}            & \multicolumn{1}{c|}{10,94}                        & \multicolumn{1}{c|}{0,32}            & \multicolumn{1}{c|}{20,72}              & 3,54                 \\ \hline
\end{tabular}%
}
\end{table}

In distributed training with 
, increasing the data volume led to a $27.37\%$ rise in network packet transmission with data augmentation (DA) and a shallow \ac{CNN}, and an $89.73\%$ increase with DA and a deep \ac{CNN}. This rise in network packet transmission is due to the need for frequent gradient synchronization across \acp{GPU}. Differences in \ac{GPU} performance, such as between the RTX 1080 and RTX 4060, can further amplify this communication load, as the faster \ac{GPU} must wait for the slower one to synchronize.

 \textbf{Limitations.} The inclusion of other data augmentation methods may change the volume of network packet transmission in distributed training, potentially affecting scalability and efficiency. Similarly, the early stopping applied during training may have restricted further insights into long-term model performance trends, limiting the comprehensive evaluation across all epochs.

\section{Concluding Remarks}\label{sec:conclusion}

This paper evaluated the impact of distributed learning on different response variables using \ac{CNN} and a computer vision dataset of rice diseases. Understanding these behavioral nuances is crucial, as it provides reliability and predictability when deploying such technologies in production environments. We observed that, while state-of-the-art research focuses on model explainability, it often overlooks the impact of these models on different response variables, especially in the context of computer vision.

Among our findings, we assessed that introducing \ac{DA} in distributed training has a significant effect on the underlying infrastructure. For future work, we are developing analyses that consider other factors, such as \ac{CNN} parameters and different datasets. This study offers valuable insights for the deployment phases of \ac{CNN} models in real-world environments, where neglecting these effects can significantly impact energy consumption and data center performance.

\section*{Acknowledgments}

This study was financed in part by the Coordenação de Aperfeiçoamento de Pessoal de Nível Superior – Brasil (CAPES) – Finance Code 001. Rodrigo Moreira gratefully acknowledges the financial support of FAPEMIG (Grant \#APQ00923-24). We also acknowledge the financial support of the FAPESP MCTIC/CGI Research project 2018/23097-3 - SFI2 - Slicing Future Internet Infrastructures.

\section*{Author contributions}
VFJ, ETM, and YSL: Software, Investigation, Validation, Visualization, Writing - original draft. FOS, LFRM, and RM: Conceptualization, Methodology, Visualization, Supervision, Funding acquisition, Resources, Writing - Review \& Editing, and Project administration.


\phantomsection

\makeatletter
\renewcommand\@biblabel[1]{{\parbox{0.7cm}{[#1]}}}
\makeatother
\renewcommand{\refname}{References}
\bibliography{template/bibtex}

\balance
\end{document}